\documentclass[11pt]{article}
\usepackage{amsmath, amssymb}
\usepackage{algorithmic}
\usepackage{algorithm}
\usepackage{graphicx}
\usepackage{color}

\def\EE{\mathbb E}

\def\RR{\mathbb R}

\def\ZZ{\mathbb Z}

\def\cA{\mathcal A}

\def\cS{\mathcal S}

\def\cT{\mathcal T}

\def\bzero{\mathbf 0}

\def\b1{\mathbf 1}

\def\eps{\varepsilon}

\newcommand{\qed}{\hfill$\square$\bigskip}
\newcommand{\raf}[1]{(\ref{#1})}
\newcommand{\proof}{\noindent {\bf Proof}~~}

\newcommand{\poly}{\operatorname{poly}}
\newcommand{\polylog}{\operatorname{polylog}}
\newcommand{\argmin}{\operatorname{argmin}}

\newcommand{\argmax}{\operatorname{argmax}}
\newcommand{\conv}{\operatorname{conv}}

\newcommand{\lambdaplus}{{\lambda^+}}
\newcommand{\hide}[1]{}

\newcommand{\set}[2]{\{ #1 | #2 \}}

\newtheorem{theorem}{Theorem}
\newtheorem{lemma}{Lemma}
\newtheorem{claim}{Claim}
\newtheorem{corollary}{Corollary}
\newtheorem{proposition}{Proposition}

\newcommand{\ignore}[1]{}

\title{On Randomized Fictitious Play for Approximating Saddle Points  Over Convex Sets}
\author{
Khaled Elbassioni\thanks {Masdar Institute of Science and Technology,
Abu Dhabi, UAE
(kelbassioni@masdar.ac.ae)}
\and
Kazuhisa Makino\thanks{Graduate
School of Information Science and Technology, University of Tokyo,
Tokyo, 113-8656, Japan; (makino@mist.i.u-tokyo.ac.jp)}
\and
Kurt Mehlhorn\thanks{Max Planck Institute for Informatics;
Campus E1 4, 66123, Saarbruecken, Germany
(mehlhorn@mpi-inf.mpg.de)}
\and
Fahimeh Ramezani\thanks{Max Planck Institute for Informatics;
Campus E1 4, 66123, Saarbruecken, Germany
(ramezani@mpi-inf.mpg.de)}
}
\begin{document}
\date{}
\maketitle

\begin{abstract}
Given two bounded convex sets $X\subseteq\RR^m$ and $Y\subseteq\RR^n,$ specified by membership oracles, and
a continuous convex-concave function $F:X\times Y\to\RR$, we consider the
problem of computing an $\eps$-approximate saddle point, that is, a pair
$(x^*,y^*)\in X\times Y$ such that $\sup_{y\in Y} F(x^*,y)\le \inf_{x\in
  X}F(x,y^*)+\eps.$ Grigoriadis and Khachiyan (1995) gave a simple randomized  variant of fictitious play for computing an $\eps$-approximate saddle point  for matrix games, that is, 
when $F$ is bilinear and the sets $X$ and $Y$ are simplices. In this paper, we
extend their method to the general case.  In particular, we show that, for functions of constant ``width'', an $\eps$-approximate saddle point can be computed using $O^*(\frac{(n+m)}{\eps^2}\ln R)$ random samples from log-concave distributions over the convex sets $X$ and $Y$.  It is assumed that $X$ and $Y$ have inscribed balls of radius $1/R$ and circumscribing balls of radius $R$. As a consequence, we obtain a simple randomized polynomial-time algorithm that computes such an approximation faster than known methods for problems with bounded width and when $\eps \in (0,1)$ is a fixed, but
arbitrarily small constant. Our main tool for achieving this result is the combination of the randomized fictitious play with the recently developed results on sampling from convex sets. 
\end{abstract}

\section{Introduction}\label{intro}
Let $X\subseteq\RR^m$ and $Y\subseteq\RR^n$ be two \emph{bounded convex} sets. We assume that each set is given by a \emph{membership oracle},
that is an algorithm which given $x\in\RR^m$ (respectively, $y\in\RR^n$) determines, in polynomial time in $m$ (respectively, $n$), whether or not $x\in X$ (respectively, $y\in Y$). 
Let $F:X\times Y\to\RR$ be a continuous convex-concave function, that is, $F(\cdot,y):X\to\RR$ is convex for all $y\in Y$ and $F(x,\cdot):Y\to\RR$ is concave for all $x\in X$.
The well-known \emph{saddle-point theorem} (see e.g. \cite{R70}) states that
\begin{equation}\label{saddle}
v^*=\inf_{x\in X}\sup_{y\in Y} F(x,y)=\sup_{y\in Y}\inf_{x\in X}F(x,y).
\end{equation}
This can be interpreted as a $2$-player zero-sum game, with one player, the
minimizer, choosing her/his strategy from a convex domain $X$, while the other
player, the maximizer, choosing her/his strategy from a convex domain $Y$. For
a pair of strategies $x\in X$ and $y\in Y$, $F(x,y)$ denotes the corresponding
payoff, which is the amount that the minimizer pays to the maximizer. An
\emph{equilibrium}, when both $X$ and $Y$ are closed, corresponds to a saddle
point, which is guaranteed to exist by \raf{saddle}, and the value of the game
is the common value $v^*$. When an approximate solution suffices or at
least one of the sets $X$ or $Y$ is open, the appropriate notion is that of
$\eps$-\emph{optimal strategies}, that a pair of strategies $(x^*,y^*)\in
X\times Y$ such that for a given desired accuracy $\eps>0$, 
\begin{equation}\label{ee}
\sup_{y\in Y} F(x^*,y)\leq\inf_{x\in X}F(x,y^*)+\eps.
\end{equation}
 
There is an extensive literature on the existence of saddle points in this
class of games and their applications, see e.g.~\cite{D63,G67,Kp00,M84,N84,R70,S58,T72,W45,Z90,B97,DKR91,W03}.  
 A particularly important case is when the sets
$X$ and $Y$ are polytopes with an exponential number of facets arising as the
convex hulls of combinatorial objects (see section \ref{App} for some applications). 

One can easily see that \raf{saddle} can be reformulated as a convex
minimization problem over a convex set given by a membership
oracle\footnote{Minimize $F(x)$, where $F(x) = \max_y F(x,y)$.}, and hence any algorithm for solving this class of problems,
e.g., the Ellipsoid method, can be used to compute a solution to \raf{ee}, in time polynomial in the input size and $\polylog(\frac{1}{\eps})$ (see, e.g., \cite{GLS93}). However, there has recently been an increasing interest in finding simpler and faster approximation algorithms for this type of problems, sacrificing the dependence on $\eps$ from $\polylog(\frac{1}{\eps})$ to $\poly(\frac{1}{\eps})$, in exchange of efficiency in terms of other input parameters; see e.g. \cite{AHK05,AK07,BBD04,GK92,GK95,GK96,GKPV01,GK98,GK04,K04,K07,LN93,KY07,Y01,DJ07,PST91}. 

In this paper, we show that it is possible to get such an algorithm for computing an $\eps$-saddle point \raf{ee}. Our algorithm is based on combining a technique developed by Grigoriadis and Khachiyan \cite{GK95}, based on a randomized variant of Brown's fictitious play \cite{B51}, with the recent results on random sampling from convex sets (see, e.g., \cite{LV06,V05}). 
Our algorithm is superior to known methods when the width parameter
  $\rho$ (to be defined later) is small and $\eps \in (0,1)$ is a fixed but
  arbitrarily small constant; see the comparison with sampling-based algorithms in Section~\ref{related}.

\section{Our Result}

We  need to make the following technical assumptions:\smallskip

(A1) We know $\xi^0\in X$, and $\eta^0\in Y$, and strictly positive numbers $r_X$, $R_X$, $r_Y$, and $R_Y$ such that $B^m(\xi^0,r_X)\subseteq X\subseteq B^m(\bzero,R_X)$ and $B^n(\eta^0,r_Y)\subseteq Y\subseteq B^n(\bzero,R_Y)$, where $B^k(x^0,r)=\{x\in\RR^k:~\|x-x^0\|_2\leq r\}$ is the $k$-dimensional ball for radius $r$ centered at $x^0\in\RR^k$. In particular, both $X$ and $Y$ are full-dimensional in their respective spaces (but maybe open). In what follows we will denote by $R$ the maximum of $\{R_X,R_Y,\frac{1}{r_X},\frac{1}{r_Y}\}$.

(A2) $|F(x,y)|\leq 1 \mbox{ for all }x\in X \mbox{ and }y\in Y.$\smallskip

Assumption (A1) is standard for algorithms that deal with convex sets defined by membership oracles (see, e.g., \cite{GLS93}), and will be required by the sampling algorithms. Assumption (A2) can be made without loss of generality, since the original game can be converted to an equivalent one satisfying (A2) by scaling the function $F$ by $\frac{1}{\rho}$, where the ``width'' parameter  is defined as $\rho=\max_{x\in X,y\in Y}|F(x,y)|$. (For instance, in case of bilinear function, i.e, $F(x,y)=x^TAy$, where $A$ is given $m\times n$ matrix and $x^T$ is the transpose of vector $x$, we have
$\rho=\max_{x\in X,y\in Y}|x^TAy|\leq \sqrt{mn}R_XR_Y\max\{|a_{ij}|~:~i\in[m],~j\in[n]\}.$)
Replacing $\eps$ by $\frac{\eps}{\rho}$, we get an algorithm that works without assumption (A2) but whose running time is proportional to $\rho^2$.
We note that such dependence on the width is unavoidable in most known algorithms that obtain $\eps$-approximate solutions and whose running time is proportional to $poly(\frac{1}{\eps})$ (see e.g. \cite{AHK06,PST91}).

We assume throughout that $\eps$ is a positive constant less than $1$.

The main contribution of this paper is to extend the randomized fictitious play
result in \cite{GK95} to the more general setting given by \raf{ee}.  
\begin{theorem}\label{main}
Assume $X$ and $Y$ satisfy assumption (A1). Then there is a randomized
algorithm that finds a pair of $\eps$-optimal strategies in an 
expected number of $O(\frac{\rho^2(n+m)}{\eps^{2}}\ln\frac{R}{\eps})$ iterations, each
computing two samples from log-concave distributions. In
particular,\footnote{Here, we apply random sampling as a black-box for each
  iteration independently; it might be possible to improve the running time if
  we utilize the fact that the distributions are slightly modified from an
  iteration to the next.}\footnote{$O^*(\cdot)$ suppresses polylogarithmic factors that depend on
  $n$, $m$ and $\eps$.}  the algorithm requires
$O^*(\frac{\rho^2(n+m)^6}{\eps^{2}}\ln{R})$ oracle
calls.
\end{theorem}
When the width is bounded and $\eps$ is a fixed constant, our algorithm needs
$O^*((n+m)^6\ln R)$ oracle calls. This is superior to known methods that compute the $\eps$-saddle point in time polynomial in $\log \frac{1}{\eps}$; see the comparison with the Ellipsoid algorithm and sampling-based algorithms in Section~\ref{related}. 
\section{Applications in combinatorial optimization}\label{App}
In this section we give  some examples for which the width parameter $\rho$ is small. 

\subsection{Mixed popular matchings}

Let $\cS,\cT$ be two families (say, of combinatorial objects), and $\cA\in[-1,1]^{\cS\times\cT}$ be a given matrix. We assume that these families have exponential size (in some input parameter) and hence, the matrix is given by an oracle that specifies for each $S\in\cS$ and $T\in\cT$ the value of $\cA(S,T)$. The objective is to find a saddle point for the matrix game defined by $\cA$ on the set of mixed strategies $\Delta_{\cS}=\{p\in\RR_+^{\cS}:~\sum_{S\in\cS}p_S=1\}$ and $\Delta_{\cT}=\{q\in\RR_+^{\cT}:~\sum_{T\in\cT}q_T=1\}$.

In general, the optimal strategies might have exponential support (i.e., an exponential number of non-zero entries). However, if the families arise from combinatorial objects in a natural way, then the supports of optimal strategies may be polynomially bounded. More precisely, let $E$ and $F$ be two sets of sizes $m$ and $n$ respectively, such that each element $S\in\cS$ (respectively, $T\in\cT$), is characterized by a vector $x(S)\in\{0,1\}^m$ indexed by the elements of $E$ (respectively, $y(T)\in\{0,1\}^n$ indexed by the elements of $F$). 
We assume further that $X=\conv\{x(S):~S\in\cS\}$ and $Y=\conv\{y(T):~T\in\cT\}$ have explicit linear descriptions, and furthermore that there exists an $m\times n$ matrix $A$ such that $\cA(S,T)=x(S)^TAy(T)$, for all $S\in\cS$ and $T\in\cT$. 
Then it follows from Von Neumann's Saddle point theorem \cite{D63} (which is a special case of \raf{saddle}) that
\begin{equation}\label{e-comb}
\min_{p\in\Delta_{\cS}}\max_{q\in\Delta_{\cT}}p^T\cA q=\min_{x\in X}\max_{y\in Y}x^TAy.
\end{equation}
Indeed, 
\begin{eqnarray*}
\min_{p\in\Delta_{\cS}}\max_{q\in\Delta_{\cT}}p^T\cA q&=&\min_{p\in\Delta_{\cS}}\max_{q\in\Delta_{\cT}}\sum_{S\in\cS,T\in\cT}p_Sq_T\cA(S,T)=\min_{p\in\Delta_{\cS}}\max_{q\in\Delta_{\cT}}\sum_{S\in\cS,T\in\cT}p_Sq_Tx(S)^TAy(T)\\
&=&\min_{p\in\Delta_{\cS}}\max_{q\in\Delta_{\cT}}\sum_{S\in\cS}p_Sx(S)^TA\sum_{T\in\cT}q_Ty(T)=\min_{p\in\Delta_{\cS}}\max_{y\in Y}\sum_{S\in\cS}p_Sx(S)^TAy\\
&=&\max_{y\in Y}\min_{p\in\Delta_{\cS}}\sum_{S\in\cS}p_Sx(S)^TAy=\max_{y\in Y}\min_{x\in X}x^TAy
=\min_{x\in X}\max_{y\in Y}x^TAy,
\end{eqnarray*}
see, e.g., \cite{KMN09}. Thus the original matrix game corresponds to a problem of the form  \raf{saddle}.

\medskip

A special case of this framework was considered in  \cite{KMN09} under the name of {\it mixed popular matchings}. Let $G=(U\cup V,E)$ be a bipartite graph, and $r:E\subset U\times V\to \ZZ$ be a 
rank function that captures  preferences of any vertex  of $U$ over the vertices in $V$  (i.e for every $(u, v_1), (u, v_2)\in E$,  $r(u, v_1)< r(u,v_2)$ if and only if 
$u$ prefers $v_1$ to $v_2$).
A $U$-matching $M:U\to V$ is an injective mapping such that $\{(u,M(u)):~u\in U\}\subseteq E$. Let $\cS=\cT=\{\{(u,M(u)):~u\in U\}:~M\mbox{ is a $U$-matching of G}\}\subseteq 2^E$.
Given $S,T\in\cS$, define $\phi(S,T)=|\{u\in U~:~r(u,S(u))<r(u,T(u))\}|/|U|$ to be the fraction of the vertices of $U$ that ``prefer'' $S$ to $T$, and $\cA(S,T)=\phi(S,T)-\phi(T,S)$.

It is well-known (see e.g. \cite{GLS93}) that the convex hull of $U$-matchings has the linear description $X=Y=\{x\in\RR_+^E:~\sum_{(u,v)\in E}x_{u,v}=1~~\forall u\in U,~\sum_{(u,v)\in E}x_{u,v}\le 1~~\forall v\in V\}$. Furthermore, if we define $A\in\RR^{E\times E}$ to be the matrix 
with entries 
$$
a_{(u,v),(u',v')}=\left\{
\begin{array}{ll}
\frac{1}{|U|} &\mbox {if $u=u'$ and $r(u,v)<r(u',v')$},\\
-\frac{1}{|U|} &\mbox {if $u=u'$ and $r(u,v)>r(u',v')$},\\
0 &\mbox{otherwise},
\end{array}
\right.
$$
then for any $S,T\in\cS$, we can write $\cA(S,T)=x(S)^TAy(T)$, where $x(S),~y(T)\in\{0,1\}^{E}$ are the characteristic vectors of $S$ and $T$, respectively. Note that in this case $\rho\leq 1$.

\medskip

Note that in the above example, the problem can be written as a linear program of \emph{polynomially} bounded size \cite{KMN09}.
However, this is not the case when the known linear descriptions of $X$ and $Y$
are not polynomially bounded, e.g., when in the above example $G$ is a general
\emph{nonbipartite} graph. In this case finding a saddle-point may require the use of the
Ellipsoid method, the sampling techniques of \cite{BV04,KV06}, or the use of
our algorithm.

\subsection{Linear relaxation for submodular set cover}
Let $f:2^{[n]}\to[0,1]$ be a {\it monotone submodular} set-function. Consider the problem of minimizing $f(X)$ subject to the constraint that the characteristic vector $e(X)\in\{0,1\}^n$ belongs to a polytope $P\subseteq\RR^n$. For instance, in the {\it submodular set covering} problem,
the polytope $P=\{x\in[0,1]^n:~\sum_{i:S_i\ni e}x_i\ge 1\text{ for all } e\in E\}$, where $S_1,\ldots,S_n\subseteq E$ are given subsets of a finite set $E$.  Let $P_f=\{y\in\RR^n_+:~\sum_{i\in X}y_i\le f(X)\text{ for all }X\subseteq[n]\}$ be the {\it polymatroid} associated with $f$. Then it is known that $f(X)=\max_{y\in P_f} e(X)^Ty$. Thus we arrive at the following saddle point computation which provides a lower bound on the optimum submodular set cover: $\min_{x\in P}\max_{y\in P_f}x^Ty$, where $\rho\le 1$. 

For other applications of polyhedral games, we refer the reader to \cite{W03}.

\section{Relation to Previous Work}\label{related}

\noindent{\bf Matrix and polyhedral games.}~ The special case when  each of the
sets $X$ and $Y$ is a polytope (or more generally, a polyhedron) and payoff  is
a bilinear function, is known as polyhedral games (see e.g. \cite{W03}). When
each of these polytopes is just a simplex we obtain the well-known class of
matrix games. Even though each polyhedral game can be reduced to a matrix game
by using the vertex representation of each polytope (see e.g. \cite{S86}), this
transformation may be (and is typically) not algorithmically efficient since
the number of vertices may be exponential in the number of facets by which each
polytope is given. 

\medskip

\noindent{\bf Fictitious play.}~
We assume for the purposes of this subsection that both sets $X$ and $Y$ are closed, and hence the infimum and supremum in \raf{saddle} are replaced 
by the minimum and maximum, respectively.
 
In \emph{fictitious play}, originally proposed by Brown \cite{B51} for matrix games, each player updates his/her strategy by applying the best response, given the opponent's current  strategy.
More precisely, the minimizer and the maximizer initialize, respectively, $x(0)=0$ and $y(0)=0$, and for $t=1,2,\ldots,$  update  $x(t)$ and $y(t)$ by
\begin{eqnarray}\label{e1}
x(t+1)&=& \frac{t}{t+1}x(t)+\frac{1}{t+1}\xi(t), \mbox{ where } \xi(t)=\argmin_{\xi\in X}F(\xi,y(t)),\\
\label{e2}
y(t+1)&=&\frac{t}{t+1}y(t)+\frac{1}{t+1}\eta(t), \mbox{ where } \eta(t)=\argmax_{\eta\in Y}F(x(t),\eta).
\end{eqnarray}
The convergence of such pair of strategies  $x^*=\lim_{t\to\infty} x(t)$, $y^*=\lim_{t\to\infty}y(t)$, for matrix games (i.e., when $X$ and $Y$ are, respectively, $m$ and $n$-dimensional \emph{simplices}, and $F(x,y)$ is a \emph{bilinear form}, that is $F(x,y)=x^TAy$, where $A$ is given $m\times n$  matrix) was established by Robinson \cite{R51}:
$v^*=F(x^*,y^*)$. Note that in this case, the best response of each player, at each step, can be chosen from the vertices of the corresponding simplex. 
A bound of $\left(\frac{2^{m+n}}{\eps}\right)^{m+n-2}$ on the time needed for convergence to an $\eps$-saddle point was obtained by Shapiro \cite{S58}. In a more recent paper, Hofbauer and Sorin \cite{HS06} showed the convergence of fictitious play for general convex-concave functions over compact convex sets.  

\medskip
  
\noindent{\bf Randomized fictitious play.}~
In \cite{GK95}, Grigoriadis and Khachiyan introduced a randomized variant of fictitious play for matrix games. Their algorithm replaces the minimum and maximum selections \raf{e1}-\raf{e2} by a smoothed version, in which, at each time step $t$, the minimizing player selects a strategy $i\in[m]$ with probability proportional to $\exp\left\{-\frac{\eps}{2}e_iAy(t)\right\}$, where $e_i$ denotes the $i$th unit vector of dimension $m$. Similarly, the maximizing player chooses strategy $j\in[n]$ with probability proportional to $\exp  \left\{\frac{\eps}{2} x(t)Ae_j\right\}$. Grigoriadis and Khachiyan proved that, if $A\in[-1,1]^{m\times n}$, then this algorithm converges, with high probability, to an $\eps$-saddle point in $O(\frac{\log (m+n)}{\eps^2})$ iterations. Each iteration takes $O(n + m)$ time.

\medskip

\noindent{\bf The multiplicative weights update method.}~ In a similar line of work, Freund and Schapire \cite{FS99} used a method, originally developed by Littlestone and Warmuth \cite{LW94}, to
give a procedure for computing $\eps$-saddle points for matrix games. Their
procedure can be thought of as a derandomization of the randomized fictitious
play described above. A number of similar algorithms have also been developed
for approximately solving special optimization problems, such as general linear
programs \cite{PST91}, multicommodity flow problems \cite{GK98}, packing and
covering linear programs \cite{PST91,GK98,GK04,KY07,Y01}, some class of convex
programs \cite{K04}, and semidefinite programs \cite{AHK05,AK07}. Arora, Hazan
and Kale \cite{AHK06} gave a meta algorithm that puts many of these results
under one umbrella. In particular, they consider the following scenario: given
a set $X$ of decisions and a finite set $Y$ of outputs, and a payoff matrix $M\in\RR^{X\times Y}$ such that $M(x,y)$ is the penalty that would be paid if decision $x\in X$ was made and output $y\in Y$ was the result, the objective is to develop a decision making strategy that tends to minimize the total payoff over many rounds of such decision making. Arora et al. \cite{AHK06,K07} show how to apply this framework to approximately computing $\max_{y\in Y}\min_{i\in[m]}f_i(y)$, given an oracle for finding $\max_{y\in Y}\sum_{i\in[m]}\lambda_i f_i(y)$ for any non-negative $\lambda\in\RR^m$ such that $\sum_{i=1}^m\lambda_i=1$, where $Y\subseteq\RR^n$ is a given convex set and $f_1,\ldots,f_m:Y\to\RR$ are given concave functions (see also \cite{K04} for similar results). 

There are two reasons why this method cannot be (directly) used to solve our
problem \raf{ee}. First, the number of decisions $m$ is infinite in our case,
and second, we do not
  assume to have access to an oracle of the type described above; we assume only a
(weakest possible) membership oracle on $Y$. Our algorithm extends the
  multiplicative update method to the computation of approximate saddle points.

\medskip
  
\noindent{\bf Hazan's Work}.~  In his Ph.D. Thesis \cite[Chapters 4 and
5]{H06}, Hazan gave an algorithm, based on multiplicative weights updates
method, for approximating the minimum of a convex function within an absolute error of $\eps$. 
This algorithm is somewhat similar to our Algorithm \ref{algo1} below, except that it chooses the point $\xi(t)\in X$, at each time step $t=1,\ldots, T$, as the (approximate) centroid of set $X$ with respect to density $p_\xi(t)=e^{\sum_{\tau=1}^{t-1}\ln(e-F(\xi,\eta(\tau)))}$, where $e$ is the base of the natural logarithm, and outputs $\frac{1}{T}\sum_{t=1}^{T} x(t)$ at the end. Theorem 4.14 in \cite{H06} suggests that a similar procedure can be used to approximate a saddle point  for convex-concave functions\footnote{
This algorithm can be written in the same form as our Algorithm \ref{algo1} below, except that it  chooses respectively the points $\xi(t)\in X$ and $\eta(t)\in Y$, at each time step $t=1,\ldots, T$, as the (approximate) centroids of the corresponding sets with respect to densities $p_\xi(t)=e^{\sum_{\tau=1}^{t-1}\ln(e-F(\xi,\eta(\tau)))}$ and $q_\eta(t)=e^{\sum_{\tau=1}^{t-1}\ln(e+F(\xi(\tau),\eta))}$ (both of which are log-concave distributions), and outputs ($\frac{1}{T}\sum_{t=1}^{T} x(t), \frac{1}{T}\sum_{t=1}^{T}y(t)$) at the end.} 
.  However, no claim was given regarding the running time or even the convergence for such an extension, and in fact, the proof technique used in Theorem 4.14 does not seem to extend to this case since the function $\ln(e-F(\xi,\eta(\tau)))$ (respectively, $\ln(e+F(\xi(\tau),\eta))$) is not concave in $\eta(\tau)$  (respectively, not convex in $\xi(\tau)$).   


\medskip

\noindent{\bf Sampling algorithms.}~ Our algorithm makes use of known
algorithms for sampling from a given log-concave distribution\footnote{that is,
  $\log f(\cdot)$ is concave} $f(\cdot)$ over a convex set $X\subseteq
\RR^m$. The currently best known result achieving this is due to Lov\'{a}sz and
Vempala (see, e.g., \cite[Theorem 2.1]{LV07}): a random walk on $X$ converges
in $O^*(\frac{m^5}{\eps^4})$ steps to a distribution within a total \emph{variation distance}
of $\eps$ from the desired exponential distribution with high probability. 

Several algorithms for convex optimization based on sampling have been recently
proposed. Bertsimas and Vempala \cite{BV04} showed how to minimize a convex
function over a convex set $X\subseteq \RR^m$, given by a membership oracle, in
time $O^*((m^5T+m^7)\log R)$, where $T$ is the time
required by a single oracle call. When the function is linear this has been
improved by Kalai and Vempala  \cite{KV06} to  $O^*(m^{4.5}T ~{\log R})$.

\medskip

Note that we can write \raf{saddle} as the convex minimization problem $\inf_{x\in X} F(x)$, where $F(x)=\sup_{y\in Y}F(x,y)$ is a convex function.  Thus, it is worth comparing the bounds  
 we obtain in Theorem \ref{main} with the bounds that one could obtain by applying the random sampling techniques
of \cite{BV04,KV06} (see Table 1 in \cite{BV04} for a comparison between these
techniques and the Ellipsoid method). Since the above program is equivalent to
$\inf\{v~:~x\in X,\mbox{ and }F(x,y)\le v\mbox{ for all }y\in Y\}$,
the
solution can be obtained by applying the technique of
\cite{BV04,KV06}, where each membership call involves another application of these
techniques (to check if $\sup_{y\in Y}F(x,y)\le v$).  The running time of the algorithm is bounded by $O^*(n^{4.5} (m^5 T + m^7) \log^{O(1)} R)$,
 which is
significantly greater \footnote{It is also worth comparing the bound in Theorem~\ref{main} with the running time of the Ellipsoid method. Under Assumption (A1), the Ellipsoid method can be used to minimize a linear function over a convex set $X\subseteq \RR^m$ given by a membership oracle in  time $O(m^{10}T\log R+m^{12}\log R)$ (see \cite{GLS93} and Table 1 in \cite{BV04}). In the special case when $F(x,y)$ is linear in $y$, this implies (by a similar argument as the one given above) a total running time of  $O^*((n^{10}(m^{10}T+m^{12})+n^{12})\log^{O(1)} R)$ which is
significantly greater than the bound stated in Theorem~\ref{main}.} than the bound stated in Theorem~\ref{main}. Note, however, that these algorithms, unlike our algorithm, depend only polylogarithmically on $\frac{1}{\epsilon}$.
  
\section{The Algorithm}
Our algorithm~\ref{our algorithm} is an adaptation of the algorithms in \cite{GK95} and \cite{FS99}. It proceeds in steps $t=0,1,\ldots$, updating the pair of \emph{accumulative} strategies $x(t)$ and $y(t)$. Given the current pair $(x(t),y(t))$, define 
\begin{eqnarray}\label{e1.1}
p_{\xi}(t)&=&e^{-\frac{\eps t F(\xi, y(t))}{2}}~~  \mbox{ for $\xi\in X$,}  \\
\label{e1.2}
q_{\eta}(t)&=&e^{\frac{\eps t F(x(t), \eta)}{2}}~~ \mbox{ for  $\eta\in Y$,}
\end{eqnarray}
and let
\begin{equation*}
\|p(t)\|_1=\int_{\xi\in X} p_{\xi}(t)d\xi  ~~~~\mbox{ and }~~~~  
\|q(t)\|_1=\int_{\eta\in Y}q_{\eta}(t)d\eta
\end{equation*}
be the respective normalization factors.
The parameter $T$ will be specified later (see Lemma \ref{l-4}).

\begin{algorithm}[t]
\caption{Randomized fictitious play}
\label{algo1}
\begin{algorithmic}[1]
\REQUIRE  Two convex bounded sets $X,Y$ and a function $F(x,y)$ such that $F(\cdot,y):X\to\RR$ is convex for all $y\in Y$ and $F(x,\cdot):Y\to\RR$ is concave for all $x\in X$, satisfying assumptions (A1) and (A2)
\ENSURE A pair of $\eps$-optimal strategies 
\STATE $t:=0$;  choose $x(0)\in X$; $y(0)\in Y$, arbitrarily
\WHILE  {$t\leq T$}
\STATE Pick $\xi\in X$ and $\eta\in Y$, independently, from $X$ and $Y$ with densities $\frac{p_{\xi}(t)}{\|p(t)\|_1}$ and $\frac{q_{\eta}(t)}{\|q(t)\|_1}$, respectively  
\label{r}
\STATE  $x(t+1):=\frac{t}{t+1}x(t)+\frac{1}{t+1}\xi$; $y(t+1):=\frac{t}{t+1}y(t)+\frac{1}{t+1}\eta$; $t:=t+1$;
\ENDWHILE
\RETURN (x(t)$, y(t)$)
 \end{algorithmic}\label{our algorithm}
\end{algorithm}
 
\section{Analysis}\label{analysis}
Following \cite{GK95}, we use a potential function $\Phi(t)=\|p(t)\|_1\|q(t)\|_1$ to bound the number of iterations required by the algorithm to reach an $\eps$-saddle point. 
The analysis is composed of three parts. The first part of the analysis is a generalization of the arguments in \cite{GK95} (and \cite{KY07}): 
we show that the potential function increases, on the average, only by a factor of $e^{O(\eps^2)}$, implying that after $t$ iterations the potential is at most a factor of $e^{O(\eps^2)t}$ of the initial potential. While this was  enough to bound the number of iterations by  $O(\eps^{-2}\log(n+m))$ when both $X$ and $Y$ are simplices and the potential is a sum over all vertices of the simplices \cite{GK95}, this cannot be directly applied in our case. This is because of the fact that a definite integral of a non-negative function over a given region $Q$ is bounded by some $\tau$ does not imply that the function at any point in $Q$ is also bounded by $\tau$. In the second part of the analysis, we overcome this difficulty by showing that, due to concavity of the exponents in \raf{e1.1} and \raf{e1.2}, the change in the function around a given point cannot be too large, and hence, the value at a given point cannot be large unless there is a sufficiently large fraction of the volume of the sets $X$ and $Y$ over which the integral is also too large.

In the last part of the analysis, we show that the same bound on the running time holds when the sampling distributions in line \ref{r} of the algorithm are replaced by sufficiently close approximate distributions.

\subsection{Bounding the potential increase}
\begin{lemma}\label{l1}
For $t=0,1,2,\ldots,$
$$
\EE[\Phi(t+1)]\leq \EE[\Phi(t)](1+\frac{\eps}{6}^2)^2.
$$
\end{lemma}
\proof
Conditional on the values of $x(t)$ and $y(t)$, we have
\begin{eqnarray}\label{e3.00}
\|p(t+1)\|_1&=&\int_{\xi\in X} e^{-\frac{\eps(t+1)F(\xi, y(t+1))}{2}} d\xi =\int_{\xi\in X} e^{-\frac{\eps(t+1)F(\xi, \frac{t}{t+1}y(t)+\frac{1}{t+1}\eta)}{2}} d\xi\nonumber\\
&\leq&\int_{\xi\in X} e^{-\frac{\eps tF(\xi, y(t))}{2}}  e^{-\frac{\eps  F(\xi, \eta)}{2}}d\xi
= \int_{\xi\in X} p_{\xi}(t) e^{-\frac{\eps F(\xi, \eta)}{2}}d\xi\nonumber\\
&&\hspace{-2cm}\leq\int_{\xi\in X} p_{\xi}(t) \left[1+\frac{\eps^2}{6}-\frac{\eps}{2}F(\xi, \eta)\right] d\xi
=\|p(t)\|_1(1+\frac{\eps^2}{6}-\frac{\eps}{2}\frac{\int_{\xi\in X} p_{\xi}(t)F(\xi,\eta)d\xi}{\|p(t)\|_1}),\nonumber
\end{eqnarray}
using assumption (A2), concavity of $F(\xi, \cdot):Y\to\RR$ and the inequality $e^\delta\leq 1+\delta+\frac{2}{3}\delta^2$, valid for all $\delta\in [-\frac{1}{2},\frac{1}{2}]$.
Taking the expectation with respect to $\eta$ (with density proportional to $q_{\eta}(t)$), we get 
\begin{equation}\label{e3.1}
\EE_{q}[\|p(t+1)\|_1]\le \|p(t)\|_1\left[1+\frac{\eps^2}{6}-\frac{\eps}{2}\frac{\int_{\eta\in Y}q_{\eta}(t) \int_{\xi\in X} p_{\xi}(t)F(\xi,\eta)d\xi d\eta}{\|q(t)\|_1\|p(t)\|_1}\right].
\end{equation} 
Similarly, by taking the expectation with respect to $\xi$ 
(with density proportional to $p_{\xi}(t)$),
we can derive

\begin{equation}\label{e3.2}
\EE_{p}[\|q(t+1)\|_1]\leq \|q(t)\|_1\left[1+\frac{\eps^2}{6}+\frac{\eps}{2}\frac{\int_{\xi\in X}p_{\xi}(t)\int_{\eta\in Y} q_{\eta}(t)F(\xi,\eta)d\eta d\xi}{\|p(t)\|_1\|q(t)\|_1}\right].
\end{equation}
Now, using independence of $\xi$ and $\eta$, we have
\begin{eqnarray*}
\EE[\Phi(t+1)|x(t),y(t)]&\le& \Phi(t)\left[\left(1+\frac{\eps^2}{6}\right)^2
\right.\\
&&\hspace{-3.5cm}\left.
+~\frac{\eps}{2}\left(1+\frac{\eps^2}{6}\right)\left(\frac{\int_{\xi\in X}p_{\xi}(t)\int_{\eta\in Y} q_{\eta}(t)F(\xi,\eta)d\eta d\xi}{\|p(t)\|_1\|q(t)\|_1}
-~\frac{\int_{\eta\in Y}q_{\eta}(t) \int_{\xi\in X} p_{\xi}(t)F(\xi,\eta)d\xi d\eta}{\|q(t)\|_1\|p(t)\|_1}\right)
\right.\\
&&\hspace{-3.5cm}\left.
- \frac{\eps^2}{4}\frac{\int_{\xi\in X}p_{\xi}(t)\int_{\eta\in Y} q_{\eta}(t)F(\xi,\eta)d\eta d\xi}{\|p(t)\|_1\|q(t)\|_1}.\frac{\int_{\eta\in Y}q_{\eta}(t) \int_{\xi\in X} p_{\xi}(t)F(\xi,\eta)d\xi d\eta}{\|q(t)\|_ 1\|p(t)\|_1}\right].
\end{eqnarray*}
By interchanging the order of integration, we get that the second part of the sum on the right-hand side is zero, and third part is non-positive.
Hence,
\begin{equation}\label{e3}
\EE[\Phi(t+1)|x(t),y(t)]\le \Phi(t)\left(1+\frac{\eps^2}{6}\right)^2.
\end{equation}
The lemma follows by taking the expectation of \raf{e3} with respect to $x(t)$ and $y(t)$.
\qed

 By Markov's inequality we have the following statement.
\begin{corollary}\label{c1}
With probability at least $\frac{1}{2}$, after $t$ iterations,
\begin{equation}\label{bd}
\Phi(t)\le 2e^{\frac{\eps^2}{3}t}\Phi(0).
\end{equation}
\end{corollary}
At this point one might be tempted to conclude the proof, as in \cite{GK95,KY07}, by implying from Corollary \ref{c1} and the non-negativity of the function under the integral
\begin{equation}\label{potential}
\Phi(t)=\int_{\xi\in X,\eta\in Y}e^{\frac{\eps}{2}t(F(x(t),\eta)-F(\xi, y(t)))}d\xi d\eta,
\end{equation}
that this function is bounded at every point also by $2e^{\frac{\eps^2}{3}t}\Phi(0)$ (with high probability). This would then imply that
the current strategies are $\eps$-optimal. However, this is not necessarily true in general and we have to modify the argument to show that, even though the value of the function at some points can be larger than the bound $2e^{\frac{\eps^2}{3}t}\Phi(0)$, the increase in this value cannot be more than an exponential (in the input description), which is still enough for the bound on the number of iterations to go through. 
\subsection{Bounding the number of iterations}
For convenience, define $Z=X\times Y$, and concave function $g_t:Z\to\RR$ given at any point $z=(\xi,\eta)\in Z$ by $g_t(\xi,\eta):=\frac{\eps}{2}t\left(F(x(t),\eta)-F(\xi,y(t))\right)$. 
Note that, by our assumptions, $Z$ is a full-dimensional bounded convex set in $\RR^N$ of volume $\Phi(0)=\mathop{vol}(X)\cdot \mathop{vol}(Y)$, where $N=n+m$. Furthermore, assumption (A2) implies that  for all $z\in Z$,  
\begin{align}\label{f121}
|g_t(z)|=|\frac{\eps}{2}t\left(F(x(t),\eta)-F(\xi,y(t))\right)|\leq \eps t.
\end{align} 

A sufficient condition for the convergence of the algorithm to an $\eps$-approximate equilibrium is provided by the following lemma.

\begin{lemma}\label{l2}
Suppose that \raf{bd} holds and there exists an $\alpha$ such that
\begin{eqnarray}
\label{e4.1}
0< \alpha<4\eps t,\\
\label{e4.2}
e^{\frac{1}{2}\alpha}\left(\frac{\alpha}{4\eps t}\right)^N\mathop{vol}(Z)>1.
\end{eqnarray}
Then 
\begin{equation}\label{e5}
e^{g_t(z)}\le 2e^{\frac{\eps^2}{3}t+\alpha}\Phi(0)\mbox{ for all $z\in Z$}.
\end{equation}
\end{lemma}
\begin{figure}[t]
\vspace{-3cm}
\begin{center}
\includegraphics[width=0.7\textwidth]{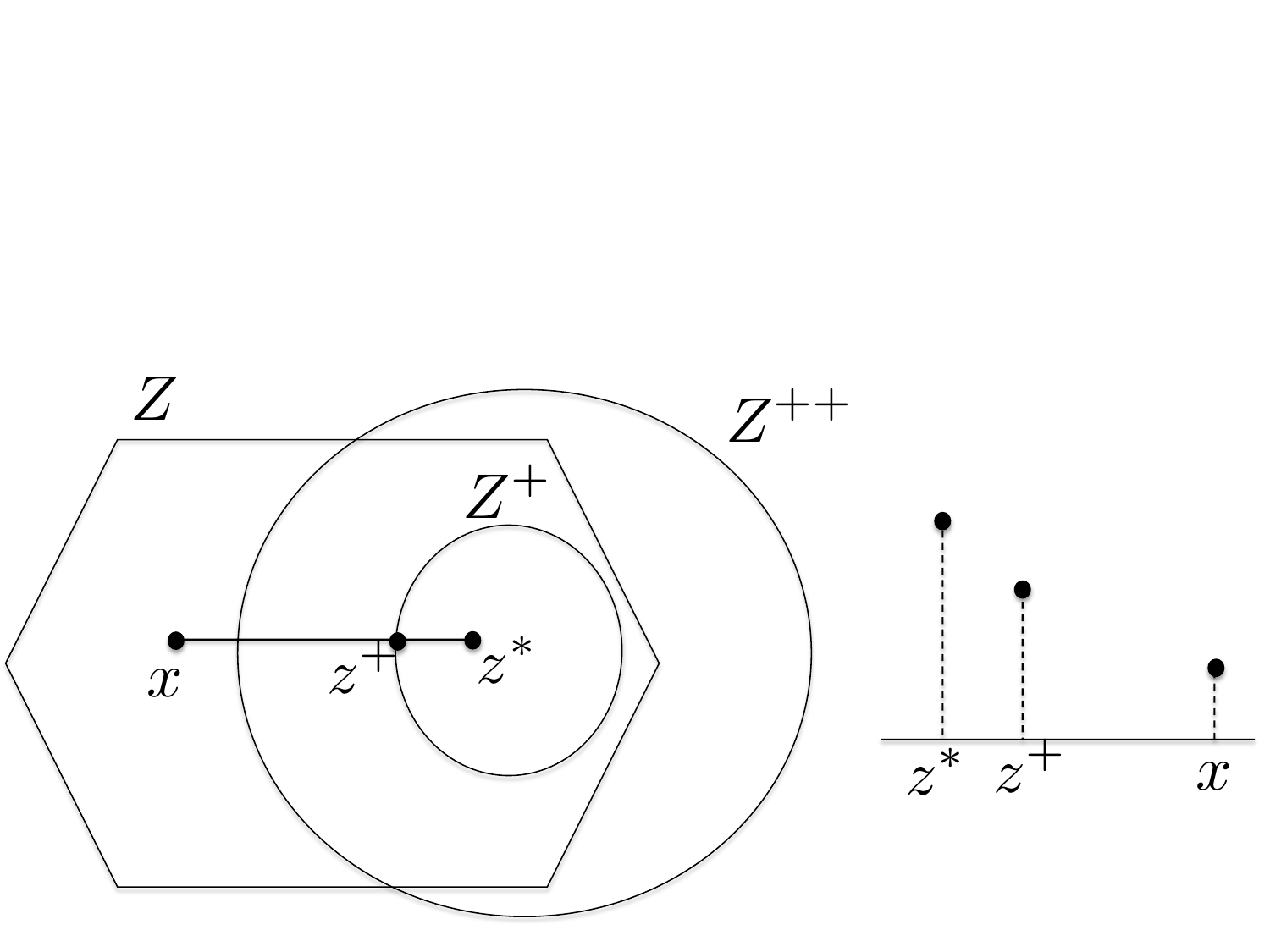}
\end{center}
\caption{\label{fig: Lemma2} The drawing on the left illustrates the notation used in the proof of Lemma~\ref{l2}. We assume for the sake of a contradiction that there is a points $x \in Z \setminus Z^{++}$. Observe that $Z^{++}$ is a scaled version of $Z^+$. The drawing on the right illustrates the contradiction. A function that drops by $\alpha/2$ from $z^*$ to $z^+$ and by at most $2 \eps t$ from $z^*$ to $x$ cannot be concave. }
\end{figure}
\proof Figure~\ref{fig: Lemma2} illustrates the definitions used in the proof of Lemma~\ref{l2}.
 Assume otherwise, i.e., there is $z^* \in Z$ with $g_t(z^*) >
  \frac{\eps^2}{3}t + \alpha + \ln(2 \Phi(0))$. Let $\lambda^* = \alpha/(4
  \eps t) < 1$, 
\[Z^+ = \{z \in Z| g_t(z) \ge  g_t(z^*) - \alpha/2\}, \text{ and }
Z^{++} = \{z^*  +\frac{1}{\lambda^*}(z - z^*)|z \in Z^+\}.\]
Concavity of $g_t$ implies convexity of $Z^+$.  Thus, for every  $z\in Z^+$ and every $\lambda'$, $0 \le \lambda' \le 1/\lambda^*$, 
we have  $\lambda^* \lambda' z +(1-\lambda^*\lambda')z^*\in Z^+$, and hence
\[  z^* + \frac{1}{\lambda*} ( \lambda^* \lambda' z +(1-\lambda^*\lambda')z^* - z^*) = z^*+\lambda'(z-z^*) \in Z^{++}. \]
Thus, for every $z \in Z^+$, the entire ray $\set{z^* + \lambda'(z - z^*)}{0 \le \lambda' \le 1/\lambda^*}$ belongs to $Z^{++}$. In particular,  $Z^+\subseteq Z^{++}$. 

We next show 
 $Z \subseteq Z^{++}$. Toward a contradiction assume that  $x \in Z \setminus Z^{++}$ (and hence $x\in Z\setminus Z^+$). Let us define
\[\lambda^+= \sup \{ \lambda~|~{ z^*  + \lambda(x - z^*) \in Z^+}\}~ 
\text{and }  z^+ = z^* + \lambda^+(x - z^*).\]
By continuity of $g_t$, $z^+ \in Z^+$ and $g_t(z^*) - \alpha/2 = g_t(z^+)$. By definition of $z^+$, we have $x - z^* = \frac{1}{\lambdaplus}
(z^+ - z^*)$ and hence
\begin{align}\label{f122}
\frac{1}{\lambdaplus} > \frac{1}{\lambda^*}.
\end{align}
 But $z^+ =
\lambda^+x + (1 - \lambdaplus) z^*$ and hence
\[ g_t(z^*) - \alpha/2 = g_t(z^+) = g_t(\lambda^+x + (1 - \lambdaplus) z^*)
\ge \lambda^+g_t(x) + (1 - \lambdaplus) g_t(z^*).\]
Thus
\[   \frac{\alpha}{2} \le \lambda^+(g_t(z^*) - g_t(x)) \le \lambda^+(|g_t(z^*)| +| g_t(x)|)\le  2 \eps t
\lambda^+\]
where the last inequality comes from \raf{f121} because $z^*,x\in Z$. Therefore we have $\lambda^+\ge \frac{\alpha}{4\eps t}=\lambda^*$ which contradicts \raf{f122}.

We have now established $Z \subseteq Z^{++}$. By definition, we have $Z^{++}=\frac{1}{\lambda^*}Z^++(1-\frac{1}{\lambda^*})z^*$. Since the volume of a body is invariant under translation, we have  
\[   \mathop{vol}(Z) \le  \mathop{vol}(Z^{++}) =\mathop{vol}\left(\frac{1}{\lambda^*}Z^{+}\right)= \left(\frac{1}{\lambda^*}\right)^N  \mathop{vol}(Z^+)\]
and further
\begin{align*}
\Phi(t) &= \int_{z \in Z} e^{g_t(z)} dz \ge \int_{z \in Z^+} e^{g_t(z)} dz \\ &\ge 2
\Phi(0)e^{\frac{\eps^2}{3} t + \frac{1}{2} \alpha}  \mathop{vol}(Z^+) 
\ge 2
\Phi(0)e^{\frac{\eps^2}{3} t + \frac{1}{2} \alpha} \left(\frac{\alpha}{4 \eps
  t}\right)^N \mathop{vol}(Z) >  2 \Phi(0)e^{\frac{\eps^2}{3} t},
\end{align*}
a contradiction to \raf{bd}.
\qed

We can now derive an upper-bound on the number of iterations needed to converge to $\eps$-optimal strategies.
\begin{lemma}\label{l-3}
If \raf{e5} holds, $\eps \in (0,1)$, $\alpha > 0$, and
\begin{equation}\label{e4.3}
t\geq \frac{6}{\eps^2}(\alpha+\max\{0,\ln (2\mathop{vol}(Z))\}),
\end{equation}
then $(x(t),y(t))$ is an $\eps$-optimal pair and \raf{e4.1} holds.
\end{lemma}
\proof
By \raf{e5} we have $g_t(z)\leq \frac{\eps^2}{3}t+\alpha+\ln(2\Phi(0))= \frac{\eps^2}{3}t+\alpha+\ln(2\mathop{vol}(Z))$ for all $z\in Z$, or equivalently,
$$\frac{\eps}{2}t(F(x(t),\eta)-F(\xi,y(t)))\leq \frac{\eps^2}{3}t+\alpha+\ln(2\mathop{vol}(Z))~~~~\text{for all}~ \xi \in X \text{~and~} \eta \in Y.$$ 
Hence,
$$F(x(t),\eta) \leq F(\xi,y(t))+\frac{2\eps}{3}+\frac{2}{\eps t}(\alpha+\ln(2\mathop{vol}(Z))~~~~\text{for all}~ \xi \in X \text{~and~} \eta \in Y,$$
which implies by \raf{e4.3} that
$$F(x(t),\eta) \leq F(\xi,y(t))+\eps~~~~\text{for all}~ \xi \in X \text{~and~}
\eta \in Y.$$
Finally, \raf{e4.1} holds since $4 \eps t \ge 24 \alpha/\eps > \alpha$. 
\qed

\begin{lemma}\label{l-4}
 For any $\eps\in(0,1)$, there exist $\alpha$ and 
\[   t= O\left( \frac{N}{\eps^2} \ln \frac{R}{\eps} \right)\]
satisfying \raf{e4.1},  \raf{e4.2} and \raf{e4.3}.
\end{lemma}
\proof
If $\mathop{vol}(Z)\leq \frac{1}{2}$.
Let us choose  $t = \frac{6 \alpha}{\eps^2}$. Then  $\raf{e4.2}$ becomes (after taking logarithms)
\[\frac{\alpha}{2} + N \ln \left(\frac{\alpha}{4\eps t}\right)+\ln (\mathop{vol}(Z)) > 0.\]
So choosing $\frac{\alpha}{2} = N \ln (\frac{25}{\eps})-\ln(\mathop{vol}(Z)))$ would satisfy this inequality. Then
\[t=O\left(\frac{N}{\eps^2} \ln \frac{1}{\eps}+\frac{1}{\eps^ 2}\ln
  \frac{1}{\mathop{vol}(Z)}\right).\]
Since $1/\mathop{vol}{Z} \le R^N$, the claim follows. 

If $\mathop{vol}(Z)> \frac{1}{2}$ then 
\[
 e^{\frac{\alpha}{2}} \left(\frac{\alpha}{4 \eps t}\right)^N \mathop{vol}(Z)> \frac{1}{2} e^{\frac{\alpha}{2}} \left(\frac{\alpha}{4 \eps t}\right)^N,
 \]
 Thus, in order to satisfy \raf{e4.2}, it is enough to find $\alpha$ and $t$ satisfying
 \[ \frac{1}{2} e^{\frac{\alpha}{2}} \left(\frac{\alpha}{4 \eps t}\right)^N>1.
 \]
 To satisfy \raf{e4.3}, let us simply choose $t= \frac{6 \alpha}{\eps^2}+  \frac{6}{\eps^2}\ln(2 \mathop{vol}(Z))$ and demand that
\[\frac{1}{2} e^{\frac{\alpha}{2}} \left(\frac{\alpha}{4 \eps t}\right)^N= \frac{1}{2} e^{\frac{\alpha}{2}} \left(\frac{\alpha}{ \frac{24 \alpha}{\eps}+ \frac{24 }{\eps}  \ln(2 \mathop{vol}(Z))}\right)^N>1,
\]
or equivalently, 
\[2\left(\frac{24}{\eps}\right)^N\left(1+\frac{ \ln(2 \mathop{vol}(Z))}{\alpha}\right)^N< e^{\frac{\alpha}{2}}. \]
Thus, it is enough to select $\alpha=  \max \left\{4(\ln2+N\ln(\frac{24}{\eps})), 2 \sqrt{N\ln(2\mathop{vol}(Z))} \right\}$ which satisfies 
\begin{eqnarray*}
2\left(\frac{24}{\eps}\right)^N \le e^{\frac{\alpha}{4}}
~\text { and }~
\left(1+\frac{\ln(2 \mathop{vol}(Z))}{\alpha}\right)^N< e^{\frac{\ln(2 \mathop{vol}(Z))}{\alpha}N}\le e^{\frac{\alpha}{4}}.
\end{eqnarray*}
It follows that
\[t =\max \left\{\frac{24}{\eps^2}(\ln2+N\ln(\frac{24}{\eps})),
  \frac{12}{\eps^2} \sqrt{N\ln(2\mathop{vol}(Z))} \right\}+
\frac{6}{\eps^2}\ln(2\mathop{vol}(Z)).\] 
Since $\mathop{vol}(Z) \le R^N$, the claim follows. 

In both cases \raf{e4.1} holds by the preceding lemma. \qed

%
%
%

\begin{corollary}\label{c3}
Assume $X$ and $Y$ satisfy assumptions (A1) and (A2). Then Algorithm \ref{algo1}, when run with $T$ satisfying the bound in Lemma \ref{l-4}, computes a pair of $\eps$-optimal strategies in
expected $O(\frac{n+m}{\eps^{2}}\ln\frac{R}{\eps})$ iterations. 
\end{corollary}

%

\subsection{Using approximate distributions}
We now consider the (realistic) situation when we can only sample approximately from the convex sets. In this case we assume the existence of approximate 
sampling routines that, upon the call in step \ref{r} of the algorithm, return vectors $\xi\in X$, and (independently) $\eta\in Y$, with densities $\hat{p}_{\xi}(t)$ and $\hat{q}_{\eta}(t)$, such that
\begin{equation}\label{as}
\sup_{X'\subseteq X}\left|\frac{\hat{p}_{X'}(t)}{\hat{p}_X(t)}-\frac{p_{X'}(t)}{p_X(t)}\right|\le \delta~~\mbox{ and }~~\sup_{Y'\subseteq Y}\left|\frac{\hat{q}_{Y'}(t)}{\hat{q}_Y(t)}-\frac{q_{Y'}(t)}{q_Y(t)}\right|\le \delta,
\end{equation}
where $\hat{p}_{X'}(t)=\int_{\xi\in X'}\hat{p}_{\xi}d\xi$ (similarly, define $p_{X'}(t),\hat{q}_{Y'}(t),q_{Y'}(t)$), and $\delta$ is a given desired accuracy. 
We next prove an approximate version of Lemma \ref{l1}.

\begin{lemma}\label{l3}
Suppose that we use approximate sampling routines with $\delta=\eps/4$ in step \ref{r} of Algorithm \ref{algo1}. Then,
for $t=0,1,2,\ldots,$ we have
$$
\EE[\Phi(t+1)]\leq \EE[\Phi(t)](1+\frac{43}{36}\eps^2).
$$
\end{lemma}
\proof
The argument up to Equation \raf{e3.00} remains the same. Taking the expectation with respect to $\eta$ (with density proportional to $\hat{q}_{\eta}(t)$), we get 
\begin{equation}
\EE_{\hat{q}}[\|p(t+1)\|_1]\le \|p(t)\|_1\left[1+\frac{\eps^2}{6}-\frac{\eps}{2}\frac{\int_{\eta\in Y}\hat{q}_{\eta}(t) \int_{\xi\in X} p_{\xi}(t)F(\xi,\eta)d\xi d\eta}{\|\hat{q}(t)\|_1\|p(t)\|_1}\right].
\end{equation} 
Similarly,
\begin{equation}
\EE_{\hat{p}}[\|q(t+1)\|_1]\leq \|q(t)\|_1\left[1+\frac{\eps^2}{6}+\frac{\eps}{2}\frac{\int_{\xi\in X}\hat{p}_{\xi}(t)\int_{\eta\in Y} q_{\eta}(t)F(\xi,\eta)d\eta d\xi}{\|\hat{p}(t)\|_1\|q(t)\|_1}\right].
\end{equation}
Thus, by independence of $\xi$ and $\eta$, we have
\begin{eqnarray*}
\EE[\Phi(t+1)|x(t),y(t)]&\le& \Phi(t)\left[\left(1+\frac{\eps^2}{6}\right)^2
\right.\\
&&\hspace{-3.5cm}+\left.
\frac{\eps}{2}\left(1+\frac{\eps^2}{6}\right)\left(\frac{\int_{\xi\in X}\hat{p}_{\xi}(t)\int_{\eta\in Y} q_{\eta}(t)F(\xi,\eta)d\eta d\xi}{|\hat{p}(t)|\|q(t)\|_1}
-\frac{\int_{\eta\in Y}\hat{q}_{\eta}(t) \int_{\xi\in X} p_{\xi}(t)F(\xi,\eta)d\xi d\eta}{\|\hat{q}(t)\|_1\|p(t)\|_1}\right)
\right.\\
&&\hspace{-3.5cm}-\left.
\frac{\eps^2}{4}\frac{\int_{\xi\in X}\hat{p}_{\xi}(t)\int_{\eta\in Y} q_{\eta}(t)F(\xi,\eta)d\eta d\xi}{\|\hat{p}(t)\|_1\|q(t)\|_1}.\frac{\int_{\eta\in Y}\hat{q}_{\eta}(t) \int_{\xi\in X} p_{\xi}(t)F(\xi,\eta)d\xi d\eta}{\|\hat{q}(t)\|_1\|p(t)\|_1}\right].
\end{eqnarray*}
We will make use of the following proposition.
\begin{proposition}\label{cl1} If we set $\delta=\eps/4$ in \raf{as}, then
\begin{equation}\label{e6-}
\left|\frac{\int_{\xi\in X}\hat{p}_{\xi}(t)\int_{\eta\in Y} q_{\eta}(t)F(\xi,\eta)d\eta d\xi}{|\hat{p}(t)|\|q(t)\|_1}
-\frac{\int_{\eta\in Y}\hat{q}_{\eta}(t) \int_{\xi\in X} p_{\xi}(t)F(\xi,\eta)d\xi d\eta}{|\hat{q}(t)\|_1\|p(t)\|_1}\right|\leq \eps.
\end{equation}
\end{proposition}
\proof
Since
$$\frac{\int_{\xi\in X}p_{\xi}(t)\int_{\eta\in Y} q_{\eta}(t)F(\xi,\eta)d\eta d\xi}{\|p(t)\|_1\|q(t)\|_1}=\frac{\int_{\eta\in Y}q_{\eta}(t) \int_{\xi\in X} p_{\xi}(t)F(\xi,\eta)d\xi d\eta}{\|q(t)\|_1\|p(t)\|_1},$$
 we can bound the L.H.S. of \raf{e6-} by
\begin{equation*}
\left|\frac{\int_{\xi\in X}p_{\xi}(t)\int_{\eta\in Y} q_{\eta}(t)F(\xi,\eta)d\eta d\xi}{\|p(t)\|_1\|q(t)\|_1}-\frac{\int_{\xi\in X}\hat{p}_{\xi}(t)\int_{\eta\in Y} q_{\eta}(t)F(\xi,\eta)d\eta d\xi}{|\hat{p}(t)|\|q(t)\|_1}\right|+
\end{equation*}
\begin{equation}\label{e7}
\left|\frac{\int_{\eta\in Y}q_{\eta}(t) \int_{\xi\in X} p_{\xi}(t)F(\xi,\eta)d\xi d\eta}{\|q(t)\|_1\|p(t)\|_1}-\frac{\int_{\eta\in Y}\hat{q}_{\eta}(t) \int_{\xi\in X} p_{\xi}(t)F(\xi,\eta)d\xi d\eta}{\|\hat{q}(t)\|_1\|p(t)\|_1} \right|.
\end{equation}

Thus it is enough to show that each term in \raf{e7} is at most $\frac{\eps}{2}$. Since the two terms are similar, we only consider the first term. 
Define $X'=\{\xi\in X~:~\frac{p_{\xi}(t)}{p_X(t)}\geq \frac{\hat{p}_{\xi}(t)}{\hat{p}_X(t)}\}$ and $X''=X\setminus X'$.
\begin{eqnarray*}
&~&~\frac{1}{\|q(t)\|_1} \left|\int_{\xi\in X}\int_{\eta\in Y} q_{\eta}(t)F(\xi,\eta)\left(\frac{p_{\xi}(t)}{\|p(t)\|_1}-\frac{\hat{p}_{\xi}(t)}{|\hat{p}(t)|}\right)d\eta d\xi\right|\\
&~&~~~~~~~~~~~~~~~~~~~~~~~\leq\frac{1}{q_{Y}(t)}\int_{\xi\in X}\int_{\eta\in Y} q_{\eta}(t)|F(\xi,\eta)|\left|\frac{p_{\xi}(t)}{p_{X}(t)}-\frac{\hat{p}_{\xi}(t)}{\hat{p}_{X}(t)}\right|d\eta d\xi\\
&~&~~~~~~~~~~~~~~~~~~~~~~~\leq\frac{1}{q_{Y}(t)}\int_{\xi\in X}\int_{\eta\in Y} q_{\eta}(t)\left|\frac{p_{\xi}(t)}{p_{X}(t)}-\frac{\hat{p}_{\xi}(t)}{\hat{p}_{X}(t)}\right|d\eta d\xi~\mbox{ (by (A2))~}\\
&~&~~~~~~~~~~~~~~~~~~~~~~~=\int_{\xi\in X}\left|\frac{p_{\xi}(t)}{p_X(t)}-\frac{\hat{p}_{\xi}(t)}{\hat{p}_X(t)}\right| d\xi\\
&~&~~~~~~~~~~~~~~~~~~~~~~~=\int_{\xi\in X'}\left(\frac{p_{\xi}(t)}{p_X(t)}-\frac{\hat{p}_{\xi}(t)}{\hat{p}_X(t)}\right)d\xi+\int_{\xi\in X''}\left(\frac{\hat{p}_{\xi}(t)}{\hat{p}_X(t)}-\frac{p_{\xi}(t)}{p_X(t)}\right)d\xi\\
&~&~~~~~~~~~~~~~~~~~~~~~~~=\left(\frac{p_{X'}(t)}{p_X(t)}-\frac{\hat{p}_{X'}(t)}{\hat{p}_X(t)}\right)+\left(\frac{\hat{p}_{X''}(t)}{\hat{p}_X(t)}-\frac{p_{X''}(t)}{p_X(t)}\right)\le \frac{\eps}{2} \mbox{ (by\raf{as})}.
\end{eqnarray*}
\qed

Proposition \ref{cl1} implies that
\begin{eqnarray*}
\EE[\Phi(t+1)|x(t),y(t)]\le \Phi(t)\left[\left(1+\frac{\eps^2}{6}\right)^2+\frac{\eps^2}{2}\left(1+\frac{\eps^2}{6}\right)+\frac{\eps^4}{4}\right]\le \Phi(t)\left(1+\frac{43}{36}\eps^2\right).
\end{eqnarray*}
The rest of the proof is as in Lemma \ref{l1}.
\qed

Combining the currently known bound on the mixing time for sampling (see \cite{LV04,LV06,LV07} and also Section \ref{related}) with the bounds on the number of iterations from Corollary \ref{c3} gives Theorem \ref{main}.
\section{Conclusion}
We showed that randomized fictitious play can be applied for computing $\eps$-saddle points of convex-concave functions over the product of two convex bounded sets. Even though our bounds were stated for general convex sets, one should note that these bounds may be improved for classes of convex sets for which faster sampling procedures could be developed. 
We believe that the method used in this paper could be useful for developing algorithms for computing approximate equilibria for other classes of games. \smallskip

\section*{Acknowledgment.} 
We are grateful to Endre Boros and Vladimir Gurvich for many valuable discussions.

\bibliographystyle{alpha}
\bibliography{convex1}


\appendix

\medskip
\end{document}